\documentstyle[aps,pre,eqsecnum]{revtex}

\input epsf

\begin{document}
%%%\draft

%\twocolumn[\hsize\textwidth\columnwidth\hsize\csname %PRETTY GALLEY 2/4
%@twocolumnfalse\endcsname                            %PRETTY GALLEY 3/4

\preprint{}

\title{Cluster size distributions in particle systems with asymmetric dynamics}

\author{O. Pulkkinen\cite{EmailOtto} and 
        J. Merikoski\cite{EmailJuha}}
        
\address{
Department of Physics, University of Jyv\"askyl\"a,  
        P.O. Box 35, FIN--40351 Jyv\"askyl\"a, Finland 
}

%\date{\today}
%\date{To appear in Physical Review E}

\maketitle

\begin{abstract}
We present exact and asymptotic results for clusters in the one-dimensional 
totally asymmetric exclusion process (TASEP) with two different dynamics. 
The expected length of the largest cluster is shown to diverge 
logarithmically with increasing system size for ordinary TASEP dynamics 
and as a logarithm divided by a double logarithm for generalized 
dynamics, where the hopping probability of a particle depends on 
the size of the cluster it belongs to. 
The connection with the asymptotic theory of extreme order statistics 
is discussed in detail. 
We also consider a related model of interface growth, where the deposited 
particles are allowed to relax to the local gravitational minimum. 
\end{abstract}


%%% 02.50.Ey Stochastic processes
%%% 05.40.-a Fluctuation phenomena, random processes
%%% 81.10.Aj Theory and models of crystal growth

%\vskip2pc]                                       %PRETTY GALLEY 4 of 4

%%%%%%%%%%%%% INTRO

\section{Introduction}

The totally antisymmetric exclusion process (TASEP) is one of the 
most studied models of nonequilibrium phenomena and has attracted 
both physicists and mathematicians. 
Many of its features, including density profiles and currents,
have been solved with periodic and open boundary 
conditions \cite{DerridaPrivman97,Liggett1,Liggett2}. 
In particular, the so-called matrix formulation has provided 
an elegant way of exploring the properties of the steady state 
and has been successfully applied to get nontrivial results 
for the so-called shock solutions and for systems with 
several species of particles \cite{Derrida93_1,Derrida93_2}. 
Recently, it has been used to solve the exact large deviation functional 
for the density profile in the case of open boundaries 
\cite{Derrida_cond-mat01}.

TASEP has been used to model free 
traffic flow \cite{Schadschneider93} and traffic jams induced by a crossing 
\cite{Nagatani93}. Traffic interpretation naturally raises a question of 
queue lengths in the system. Although this problem is equivalent to 
one-dimensional site percolation in the canonical ensemble, 
if the steady state with usual exclusion dynamics and  periodic boundary 
conditions is considered, being exactly solvable, it may still 
give some insight to problems, which 
are too complicated to solve in more complex situations. 
The statistics of the longest queue, or particle cluster as 
we shall call them, is 
interesting because it measures the size of the large deviations in 
the system. Largest clusters in percolation in general dimension 
were recently studied by Bazant \cite{Bazant00}. An exact solution, 
with limiting behaviors, is presented in this article for the 
one-dimensional exclusion process. In exclusion processes, one 
has also the possibility of changing the microscopic dynamics, which 
leads to percolation problems with different types of cluster statistics. 
For example, a model for 'reckless drivers', who are trying to escape 
from their chasers is considered in the present article. 
We show that this dynamics results in Poisson statistics for 
the cluster sizes, which in turn makes the properties of the 
largest cluster very different from those in the case of 
the usual TASEP dynamics.

TASEP is also related to two other well-known nonequilibrium systems, namely
interface growth and directed polymers in random media \cite{DerridaPrivman97}.
In particular, the cluster properties are translated to the excursion 
properties of the  growing interface and the largest clusters measure 
the long excursions before turning back.

This article is organized as follows.
In Sec.~\ref{SecSystem}, we present the basic definitions 
and the mapping between TASEP and the so-called zero-range process. 
The properties of clusters in stationary TASEP are studied 
using this mapping in Sec.~\ref{SecTASEP}. In particular, we present 
analytical results for the largest particle and hole 
clusters in Secs.~\ref{SecASEPEMaxX} 
and \ref{SecFisher-Tippett}. In Sec.~\ref{SecASY} extremal statistics 
of clusters for TASEP with the modified dynamics is considered. 
Finally, in Sec.~\ref{SecDisc} we discuss a 
related model of interface growth.

\section{THE MODEL}
\protect\label{SecSystem}

The model system we analyze in Secs.~\ref{SecTASEP} and ~\ref{SecASY}
is the totally asymmetric 
exclusion process (TASEP) with $n$ particles hopping to the right 
in a one-dimensional lattice of $N$ sites, where $N$ is an even number, 
with periodic boundary conditions. 
For convenience (see below), we first restrict 
\begin{equation}
\label{rhorestricted}
\rho := n/N = 1 - 1/k, {\rm \ \ where\ \ } $k=2,3,\ldots$\ .
\end{equation}
The dynamics, however, differs from that of the ordinary TASEP. 
We introduce another parameter $z = 0,1, \ldots, n$, which for $z \geq 1$ 
makes the hopping probability to depend on the length of the queue 
behind the particle in question as follows: After randomly picking an 
occupied site, check how many occupied sites there are in between the 
chosen site and the next empty site. 
If this number is less than or equal to $z$, the rightmost 
particle in the cluster makes a move to the right - otherwise nothing happens.

Now the case $z=0$ is clearly the usual TASEP.    
In the present article, we concentrate mainly on the cluster length 
properties of the two cases $z=0$ and $z = n$. 
We shall also comment briefly on a model of interface growth, where 
the deposited particles always relax to the local minimum, 
which is equivalent to a certain symmetrization of the $z = n$ case.

\subsection{TASEP and the zero-range process}
\label{Mass-TASEP}

Many properties of the exclusion process can be accessed by using 
the mapping \cite{Evans00,Majumdar99}
to the zero-range process introduced in Ref.~\cite{Spitzer70}. 
The mass variable $M_i$, $i=1,2,...N-n$, of the zero-range process 
is the number of particles between $i^{\mathrm{th}}$ and 
$(i+1)^{\mathrm{th}}$ hole in TASEP. 
Since in TASEP there are $n$ particles and $N-n$ holes on a ring of $N$ sites, 
there will be $N-n$ mass variables each attaining integer values from zero 
to $n$ with the constraint that $\sum_{i=1}^{N-n} M_i = n$. 
A jump of a particle in TASEP corresponds to a change 
$(M_i, M_{i+1}) \rightarrow (M_i-1, M_{i+1} +1)$ for some $i$.
This way TASEP is reduced to the problem with $n$ particles hopping to 
the right on a lattice of $N-n$ sites, now with multiple occupancy allowed.

This mapping is, however, not one-to-one, because 
there are $N/(N-n)$ TASEP configuration corresponding 
to one configuration of the mass variables. 
For example, the number of all 
mass variable configurations equals the number of solutions of the equation
$ M_1 +M_2 +\ldots +M_{N-n} = n $
in non-negative integers. 
This number is ${N-1 \choose N-n-1}$. 
But ${\frac{N}{N-n}} {N-1 \choose N-n-1} = {N \choose n}$, 
which equals the number of all corresponding TASEP configurations, 
i.e.\ the number of random walk paths from $(0,0)$ to $(N,N-2n)$. 
Note that without the restriction of Eq.~(\ref{rhorestricted}) 
this relation between the numbers of configurations 
would have applied on average only.

As discussed in Ref.~\cite{Evans00}, the joint stationary distribution 
of the zero-range process corresponding to TASEP with general $z$ is 
the product of the marginal measures of the individual sites. 
For the cases $z=0$ and $z=n$, however, the stationary state is 
particularly simple and can be constructed in a straightforward way.

\section{SIMPLE EXCLUSION PROCESS, \lowercase{$z=0$}}
\label{SecTASEP}
   
\subsection{Stationary cluster size 
\label{SecASEPEX}}

It is general knowledge that the stationary distribution of TASEP on a ring 
of $N$ sites is the uniform distribution {\cite{DerridaPrivman97}}, 
the probability of each distinct configuration being ${N \choose n}^{-1}$. 
This follows immediately from the fact that the transition probability to a 
given state equals the transition probability from the state in question 
to other states, which makes the transition matrix doubly 
stochastic \cite{Bremaud99}. 
The mapping from TASEP to the zero-range process being one-to-one up to a 
multiplicative constant, the stationary distribution of the 
zero-range process corresponding to $z=0$ is also uniform.

Due to the translational invariance of our model system the mass 
variables $M_i$ are identically distributed and the general behavior 
is described by local expectations. 
The stationary distribution for size $X_i$ of 
the $i^{\rm th}$ cluster in TASEP can then be expressed in terms of the 
stationary mass variable distribution by 
\begin{equation}
\label{PXKPZ}
P(X_i = k) = P(M_i =k \vert M_i \geq 1).
\end{equation}
Next we use translational invariance and the uniformity of measures 
to conclude that Eq.~(\ref{PXKPZ}) equals the number of solutions to 
\begin{equation}
\label{M_N-1}
M_1 + M_2 + \ldots +M_{N-n-1} = n - k,
\end{equation}
where $M_i$'s are non-negative integers and $k \geq 1$, divided by the 
difference of ${N-1 \choose n}$ and the number of solutions to 
Eq.~(\ref{M_N-1}) with $k=0$. 
Therefore 
\begin{equation}
\label{ProbX_i=k}
P(X_i = k) = 
{N-k-2 \choose n-k} {N-2 \choose n-1}^{-1}.
\end{equation}
Note that one could have derived this result also by simply counting the 
random walk paths starting with exactly $k$ steps up and ending 
with a down step and normalizing this number by the number of 
random walk paths with the first step up and the last step down. 
From Eq.~(\ref{ProbX_i=k}) we immediately obtain the 
limiting distribution for the cluster lengths, 
\begin{equation}
\label{Geometric}
P_{\infty}(X_i = k) = (1-\rho)\rho^{k-1} , 
\end{equation} 
in agreement with the fact {\cite{DerridaPrivman97}} that in an 
infinite system size the measure for TASEP is the product measure.

The expected cluster size is
\begin{eqnarray}
\label{EXKPZ}
\langle X_i \rangle &=& {N-2 \choose n-1}^{-1} \sum_{k=1}^N\, k \cdot
{N-k-2 \choose n-k} \nonumber\\
&=& {\frac{1-1/N}{1-\rho}} \longrightarrow  {\frac{1}{1-\rho}}, 
\qquad {\textrm{as}}\  N\to \infty, 
\end{eqnarray}
and its variance is 
\begin{eqnarray} 
{\textrm{Var}}X_i &=& {\frac{(\rho-1/N)(1-\rho-1/N)(1-1/N)}
{(1-\rho)^2 (1-\rho+1/N)}} \nonumber\\
& &\longrightarrow {\frac{\rho}
{(1-\rho)^2}}, \qquad  {\textrm{as}}\ N\to \infty.
\end{eqnarray}

The remarkably simple result of Eq.~(\ref{EXKPZ}) shows 
that the expected cluster 
size converges to a finite value and it is the {\it number} 
of clusters that has to diverge in the $N\to \infty$ limit. 
This can be also verified  by the following direct evaluation. 
Let us first calculate the probability 
that there are $k$ clusters in the system, i.e. that 
there are $k$ occupied sites in the mass picture. If the first 
site is to be occupied, place $k$ sites on a ring, 
label one of them $1$ and fill them with $n$ identical objects, such that 
every site gets at least one object, in 
${n-1 \choose k-1}$ ways. After that, place $N-n-k$ vacant sites in between 
the $k$ occupied sites in ${N-n-1 \choose k-1}$ ways. On the other hand, 
if the site $1$ is to be vacant, place $k+1$ sites on a ring, label one 
those $1$ and fill the sites $2,3, \ldots, k$, such that every site gets at 
least one object, in ${n-1 \choose k-1}$ ways. 
Then add $N-n-k-1$ vacant sites in between in ${N-n-1 \choose k}$ ways. 
Thus we get the hypergeometric distribution,
\begin{eqnarray}
P(\# C = k) &=& {{{n-1 \choose k-1} \Big\lbrack 
{N-n-1 \choose k-1} + {N-n-1 \choose k}\Big\rbrack}\over{N-1 \choose n}}
\nonumber\\
&=& {{{n-1 \choose k-1}{N-n \choose k} }\over{N-1 \choose n}},
\end{eqnarray}
which immediately yields
\begin{equation}
\label{EnCKPZ}
\langle \# C \rangle = n{\frac{1-\rho}{1-1/N}}.
\end{equation}
From Eqs.~(\ref{EXKPZ}) and (\ref{EnCKPZ}) we find the intuitively 
expected relation $\langle \# C\rangle \cdot \langle X_i \rangle = n$.

\subsection{Length of the longest cluster} 
\label{SecASEPEMaxX}

The knowledge of expectations of maximal objects is 
of great importance since it measures the size of large deviations 
in the system. In this section we study the statistics of the longest 
cluster, $\max_i X_i$, in TASEP. In particular, we discuss the finite size 
and limiting distribution functions, related expectations and the effects 
due to discreteness of the sample space of the mass variables.

The cumulative distribution function (c.d.f.) 
\begin{equation}
F_{\rho,N}(k):=
P(\max_i X_i \leq k) = 
P(\max_{1 \leq i \leq N-n} M_i \leq k )
\end{equation} 
can be calculated as the ratio of the number of solutions to 
\begin{equation} 
\label{M_N}
M_1 + M_2 + \ldots +M_{N-n}= n
\end{equation}
in non-negative integers with $M_i \leq k$ and the number of its 
solutions with $M_i \leq n$, $\forall i$, respectively. 
The first of these two numbers equals the coefficient of 
the $n^{\rm th}$ order term in
\begin{equation}
\left( 1 + x + \ldots +x^k \right)^{N-n} = \left( {\frac{1 - x^{k+1}}{1-x}} 
\right)^{N-n}
\end{equation}
and therefore
\begin{equation}
\label{KPZCDF}
F_{\rho,N}(k) ={\frac{1}{{N-1 \choose n}}} \sum_{i=0}^{\mu_{n,N}(k)} (-1)^i 
{N-n \choose i}{N-(k+1)i-1 \choose N-n-1},
\end{equation}
where 
\begin{equation}
\label{mu}
\mu_{n,N}(k):=\min \lbrace N-n , \lfloor n/(k+1) \rfloor \rbrace
\end{equation} 
and $\lfloor \cdot \rfloor$ stands for the integer part. Note that for 
$k \geq n/2 $ the distribution function takes the simple form
\begin{equation}
\label{KPZCDFTail}
F_{\rho,N}(k) = 1 - \left( N-n \right) {\frac{{N-k-2 \choose N-n-1}}
{{N-1 \choose n}}}.
\end{equation}

The expectation calculated in terms 
of the c.d.f.\ is
\begin{eqnarray}
\label{ASEPEMaxX}
\langle\max_i X_i\rangle &=& \sum_{k=1}^n k \cdot P
(\lbrace M_i \leq k\ \forall i \rbrace 
\cap \lbrace \exists j: M_j =k\rbrace) 
\nonumber\\
&=& \sum_{k=1}^{n}\Big \lbrack 1 - P(M_i \leq k-1\ \forall i)
\Big\rbrack, 
\end{eqnarray}
so that, after simplifications, it can be written as
\begin{equation}
\label{ASEPEMaxXact}
\langle\max_i X_i\rangle ={\frac{1}{{N-1 \choose n}}} \sum_{k=1}^n 
\sum_{i=1}^{\mu_{n,N}(k-1)} (-1)^{i+1} 
{N-n \choose i}{N-ki-1 \choose N-n-1},
\end{equation}
where $\mu_{n,N}(k)$ was defined in Eq.~(\ref{mu}).

\subsection{The limiting extremal distribution}   
\label{SecFisher-Tippett}

It seems difficult to extract the limiting distribution and the type of 
the divergence of the mean from the exact results 
of Eqs.~(\ref{KPZCDF}) and (\ref{ASEPEMaxXact}) directly. 
For that purpose, we next consider an alternative approach 
based on independence.

Intuitively one would expect that as the system gets larger the weaker  
become the correlations between the clusters. In fact, it is easy to see that 
the mass variables are asymptotically independent, in that, for $k$ fixed,
\begin{equation}
\label{AsymptoticIID}
P(M_{i_1} = a_1, M_{i_2} = a_2,\ldots, M_{i_k} =a_k) \longrightarrow 
\prod_{l=1}^kP_{\infty}(M_{i_l} = a_l),
\end{equation}
where $i_l \neq i_m$, $\forall l,m$, and $N \to \infty$.
Therefore, we can approximate the c.d.f.\ $F_{\rho,N}(k)$ by using the 
product measure, i.e.
\begin{equation}
P(\max_i X_i \leq k) \approx 
P(M_1 \leq k)^{N-n}.
\end{equation}
By writing
\begin{eqnarray}
P(M_1 \leq k) &=& 1 - {N-k-2 \choose n-k-1} {N-1 \choose n}^{-1} 
\nonumber\\
&\longrightarrow& 1- \rho^{k+1}
\ \ {\textrm{as}}\ N \to \infty,
\end{eqnarray}  
we get in the large $N$ limit with $\rho$ fixed 
\begin{eqnarray}
\label{KPZCDFLim}
F_{\rho , N}(k) &\sim& \left( 1- e^{-(k+1)\log 1/\rho} \right)^{N(1-\rho)} 
\nonumber\\
&\approx& e^{-e^{-k\log 1/\rho +\log N\rho(1-\rho)}}.
\end{eqnarray}
This result is, for $k$ sufficiently large, 
in agreement with the earlier result for the tail because 
Eq.~(\ref{KPZCDFTail}) yields in the large system size limit  
\begin{equation}
F_{\rho , N}(k) \sim 1 - N\left( 1-\rho \right) e^{-(k+1)\log 1/\rho},
\end{equation}
which is simply the first order approximation for 
Eq.~(\ref{KPZCDFLim}).

The c.d.f.\ in Eq.~(\ref{KPZCDFLim}) can now be expressed in terms of a 
scaled variable 
\begin{equation}
\label{scaledvariable}
Z_{\rho,N} :=  { \frac{ \max_{i} X_i - 
{\frac{\log N\rho(1-\rho)}{\log 1/\rho}} }
{ \left(\log 1/\rho \right)^{-1} }}.
\end{equation}
From the asymptotic theory of the extremes for 
independent and identically distributed 
(i.i.d.) variables \cite{{Galambos78},{KotzKirja}} 
we then obtain the Gumbel, or Fisher-Tippett, distribution 
\begin{equation}
\label{KPZF-T}
P \left( Z_{\rho,N} \leq y \right) 
\sim e^{-e^{-y}}.
\end{equation} 
Strictly speaking, however, 
the function on the right hand side 
is continuous, whereas the c.d.f.\ on the left must be piecewise 
constant with an intrinsic 
thickness $\log1/\rho$, so that the relation holds only when 
$\left(y +\log N\rho(1-\rho)\right)/\log{(1/\rho)}$ is an integer.
In fact, Eq.~(\ref{KPZF-T}) would be the correct result for 
the {\it continuous} 
exponential parent distribution, whereas it has been shown by Anderson 
{\cite{Anderson70}} that the {\it discrete} geometric 
distribution does not belong to the domain of attraction of the 
Gumbel distribution at all, and that Eq.~(\ref{KPZF-T}) 
should indeed be replaced by 
\begin{eqnarray}
\label{limsup}
\limsup_{N\to \infty} P \left( Z_{\rho, N} \leq y \right) 
&\leq& e^{-e^{-y}}, \\
\label{liminf}
\liminf_{N\to \infty} P \left(  Z_{\rho, N} \leq y \right) 
&\geq& e^{-e^{-y + \log 1/\rho}},
\end{eqnarray}
i.e.\ the limiting distribution function has Gumbel envelopes.

\begin{figure}[htb]
\epsfxsize=\columnwidth\epsfbox{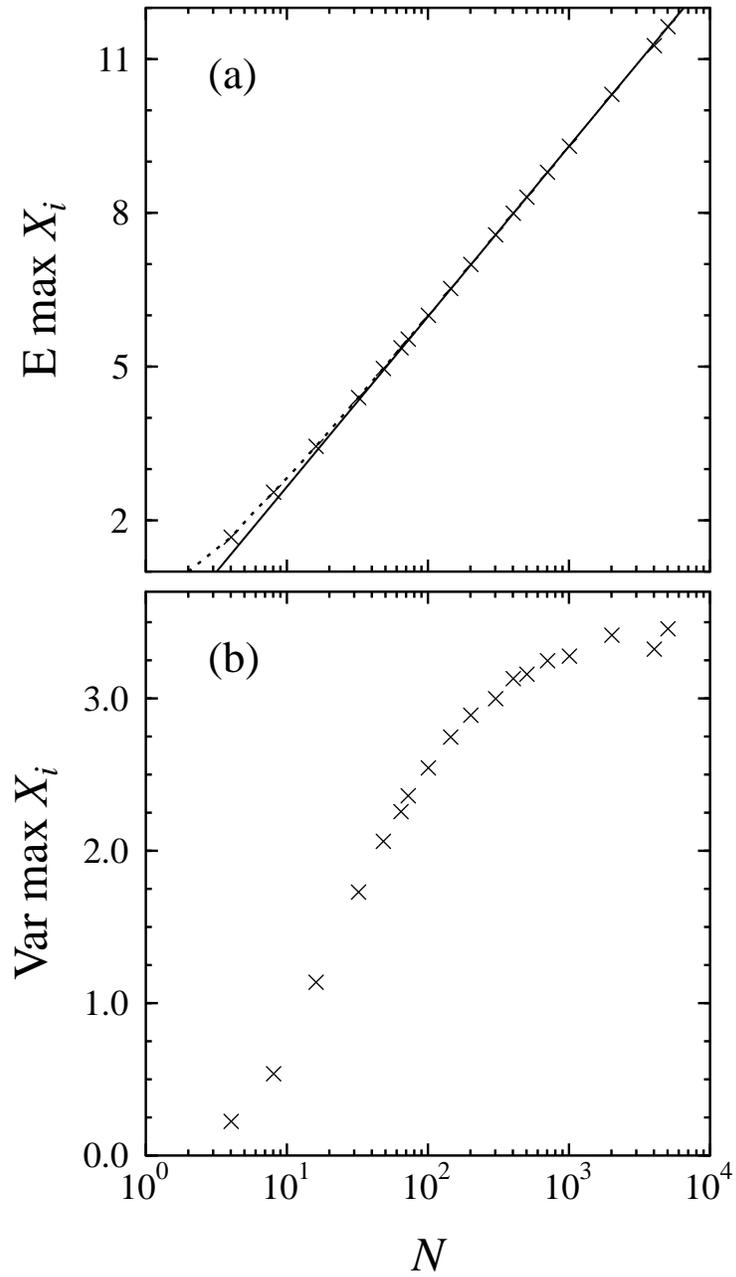}
        \caption{
          (a) Expectation value of the size of the longest cluster and 
          (b) its variance as a function of the system size $N$ 
          for $z=0$ with $\rho=1/2$. 
          In (a), the dotted curve is the exact result 
          Eq.~(\protect\ref{ASEPEMaxXact}) 
          and the full curve the asymptotic formula 
          Eq.~(\protect\ref{ASEPEMaxIID}).
          Crosses denote the simulation data. 
        \protect\label{FigASEP}}
\end{figure}

The expectation of a random variable with the Gumbel distribution is 
known to equal Euler's constant $\gamma \approx 0.5772$ 
\cite{KotzKirja}, but the evaluation of the expectation 
of a discrete variable with Gumbel envelopes is not easily found in the 
literature. A direct calculation using the proper point probability function 
seems tedious and, therefore, we argue on heuristic grounds 
that since the piecewise 
constant c.d.f.\ of the scaled variable $Z_{\rho,N}$ is 
confined between two envelope functions with expectations $\gamma$ and 
$\gamma + \log 1/\rho$, the final result should be written as 
\begin{equation}
\label{ASEPEMaxIID}
\langle\max_i X_i\rangle \approx 
{\frac{\gamma + \log N\rho(1-\rho)}{\log 1/\rho}} + {\frac{1}{2}},
\end{equation}
where $N$ is large. Fig.~\ref{FigASEP}(a) shows that the 
exact result Eq.~(\ref{ASEPEMaxXact}) and the asymptotic formula 
Eq.~(\ref{ASEPEMaxIID}) are in excellent agreement with 
the simulation data \cite{Simulation}.

The effect of discreteness of the mass variables can be seen even more 
clearly in the behavior of the variance than of the expectation. 
For continuous exponential variables, the variance of the 
maximum converges to a constant $\sigma^2 \pi^2 /6$ \cite{KotzKirja}, 
where the scaling parameter $\sigma$ equals $(\log 1/\rho)^{-1}$ in our case. 
The simulation results of Fig.~\ref{FigASEP}(b) display 
persistent fluctuations around this value, 
with the distance between the maxima diverging logarithmically in $N$.
This kind of behavior has recently been observed also by 
Bazant \cite{Bazant00} in connection with percolation 
theory and the limiting distribution functions described by 
Eqs.~(\ref{limsup}) and (\ref{liminf}) were
called Fisher-Tippett limit cycles because of the 
quasiperiodic fluctuations of the variance.

\section{TASEP WITH ASYMMETRIC DYNAMICS, \lowercase{$z=n$}}
\protect\label{SecASY}

We now turn to the case $z = n$. This means that 
every time we pick a particle at random, the rightmost particle in the 
cluster, where the chosen particle belongs to, makes a move. 
In the mass picture this translates to $n$ independent random 
walkers on a ring of $N-n$ sites \cite{Evans00}.

\subsection{Stationary cluster size}

The joint stationary distribution for $n$ independent random walkers on a 
ring of $N-n$ sites is 
\begin{equation}
P(M_1 = m_1 , \ldots , M_{N-n} =m_{N-n}) =  
{\frac{1}{(N-n)^n}} {n \choose {m_1\ \ldots \ m_{N-n}}},
\end{equation}
where $M_i$ is the number of walkers on the site $i$. 
By Eq.~(\ref{PXKPZ}), the cluster length distribution for the 
TASEP variables $X_i$ is now
\begin{equation}
P(X_i = k) =
{\frac{\left( 1- {\frac{1}{N(1-\rho)}} \right)^{\rho N -k} N^{-k} 
(1-\rho)^{-k}}
{1- \left( 1- {\frac{1}{N(1-\rho)}} \right)^{\rho N}}} { {N\rho} \choose k},
\end{equation}
the limit of which is Poisson with the proper normalization,
\begin{equation}
P_{\infty}(X_i = k) = {\frac{\lambda^{-k}}
{k! \left(e^{\lambda} - 1 \right)}}, 
\qquad {\textrm{where}}\ \ \lambda= \frac{\rho}{1-\rho}. 
\end{equation}
The parameter $\lambda$ introduced above 
describes the length scale of correlations in the system. 
Remember that in the case $z=0$ the corresponding parameter was 
$\left(\log 1/\rho \right)^{-1} \geq \lambda$, 
so that the correlations are expected to die out faster in the present case. 
Now we have the mean and the variance of the length of {\it particle} 
clusters
\begin{eqnarray}
\langle X_i \rangle &=& {\frac{\lambda}{1- \left( 1- {\frac{1}{N(1-\rho)}} 
\right)^{\rho N}}} \nonumber\\
&&\longrightarrow {\frac{\lambda}
{1 - e^{-\lambda}}} \qquad {\textrm{as}}\ N \to \infty,
\end{eqnarray}
\begin{eqnarray}
{\textrm{Var}}X_i  &=& \langle X_i \rangle 
\left( 1+ \lambda - \langle X_i \rangle\right) \nonumber\\
&&\longrightarrow {\frac{\lambda}{1 - e^{-\lambda}}} 
\left( 1+ \lambda - {\frac{\lambda}{1 - e^{-\lambda}}} \right)  
\qquad {\textrm{as}}\ N \to \infty.
\end{eqnarray} 
Since there are ${N-n \choose k}$ ways to choose $k$ sites from a ring of 
$N-n$ sites and $k!S(n,k)$, where $S(n,k)$ is the Stirling number of 
the second kind, ways to put $n$ particles into $k$ sites such that each 
site gets at least one particle, the distribution for the number of 
clusters now reads 
\begin{equation}
P(\# C = k) = {N-n \choose k} {\frac{k! S(n,k)}{(N-n)^n}}
\end{equation}
which again gives $\langle \# C \rangle \cdot \langle X_i \rangle = n$.

Due to the lack of particle-hole symmetry, 
the properties of the {\it hole} clusters differ from those presented above. 
However, the distribution for the length $H_i$ of the $i^{\mathrm{th}}$ 
hole cluster can be 
easily constructed in terms of the distribution for the mass variables,
\begin{eqnarray}
\label{HoleDistr}
P(H_i =k) &=& 
\cases{P(M_{i+1} > 0 ) & {\textrm{if $k=1$}} \cr
       P(M_{i+1} = 0,M_{i+2} = 0,\ldots, M_{i+k-1}= 0, M_{i+k} > 0) 
        & {\textrm{if $k \geq 2$}} }
\nonumber\\
&=& \left( 1 - {\frac{k-1}{N(1-\rho)}} \right)^{\rho N} - 
\left( 1 - {\frac{k}{N(1-\rho)}} \right)^{\rho N} \\
& &\longrightarrow \left( e^{\lambda} -1 \right) e^{-\lambda k} \qquad 
{\textrm{as }} N \to \infty \nonumber. 
\end{eqnarray}
The corresponding c.d.f\ is
\begin{equation}
\label{HoleCDF}
P(H_i \leq k) = 1 - 
\left( 1 - {\frac{k}{N(1-\rho)}} \right)^{\rho N} 
\longrightarrow 1 -  e^{-\lambda k} \qquad 
{\textrm{as }} N \to \infty,
\end{equation}
and the expected length of hole clusters 
\begin{equation}
\label{EH}
\langle H_i \rangle = \sum_{k=1}^{N-n} 
\left( 1 - {\frac{k-1}{N(1-\rho)}} \right)^{\rho N} 
\longrightarrow  {\frac{1}{1 -  e^{-\lambda}}} \qquad 
{\textrm{as }} N \to \infty.
\end{equation}

\subsection{Length of the longest cluster} 
\label{ASYEMaxX_i}

The exact longest particle cluster c.d.f.\ for the case $z=n$ reads
\begin{equation}
\label{ASYCDFXact} 
P(\max_i X_i \leq k) = {\frac{1}{(N-n)^n}}\sum_{{m_i \leq k,
\ \forall i,}\atop{\sum_i m_i =n}}{n \choose {m_1\ \ldots \ m_{N-n}}},
\end{equation}
which seems to be, from the viewpoint of applications, even less instructive 
than Eq.~(\ref{KPZCDF}) was for the ordinary TASEP. 
However, an approximate form for the tail of the distribution can 
be obtained from Eq.~(\ref{ASYCDFXact}) assuming that, 
for $k$ sufficiently large, at most one of the masses exceeds $k$,
\begin{eqnarray}
P(\exists j: X_j > k) &=& {\frac{1}{(N-n)^n}}\sum_{{\exists j:\, 
m_j > k,}\atop{\sum_i m_i =n}}{n \choose {m_1\ \ldots \ m_{N-n}}} \nonumber\\
&\approx& (N-n)\sum_{m_{N-n}=k+1}^{n} 
{n \choose m_{N-n}} {\frac{\left( N-n-1 \right)^{n-m_{N-n}}}
{ \left( N-n \right)^n }} 
\\ & &\times  \!\!\!
\sum_{{m_i \leq k,\ \forall i,}\atop{\sum_i m_i =n-m_{N-n}}}
{n-m_{N-n} \choose {m_1\ \ldots \ m_{N-n-1}}} {\frac{1}{(N-n-1)^{n-m_{N-n}}}}. 
\nonumber
\end{eqnarray} 
Here the sum in the last term is the probability that on a ring of 
$N-n-1$ sites with a total of $n-m_{N-n}$ independent random walkers there are 
at most $k$ walkers on a single site. But for $k \geq (n-1)/2$, and 
approximately perhaps even for smaller $k$ values, 
this probability equals unity and therefore we have
\begin{eqnarray}
\label{ASYCDFTail}
P(\max_i X_i \leq k) &\approx& 
1 - \sum_{m_{N-n}=k+1}^{n} 
{n \choose m_{N-n}} {\frac{\left( N-n-1 \right)^{n-m_{N-n}}}
{ \left( N-n \right)^{n-1} }}
\nonumber\\
&\sim& 1- N(1-\rho)\lambda^{k} e^{-\lambda} \sum_{j\geq 0}
{\frac{ \lambda^j}{j!(k+j+1)!}} \nonumber\\
&\approx& 1- n {\frac{\lambda^{k} e^{-\lambda}}{ (k+1)!}},
\end{eqnarray} 
when $k$ is sufficiently large and also $N$ large at the last two stages.

\begin{figure}[htb]
\epsfxsize=\columnwidth\epsfbox{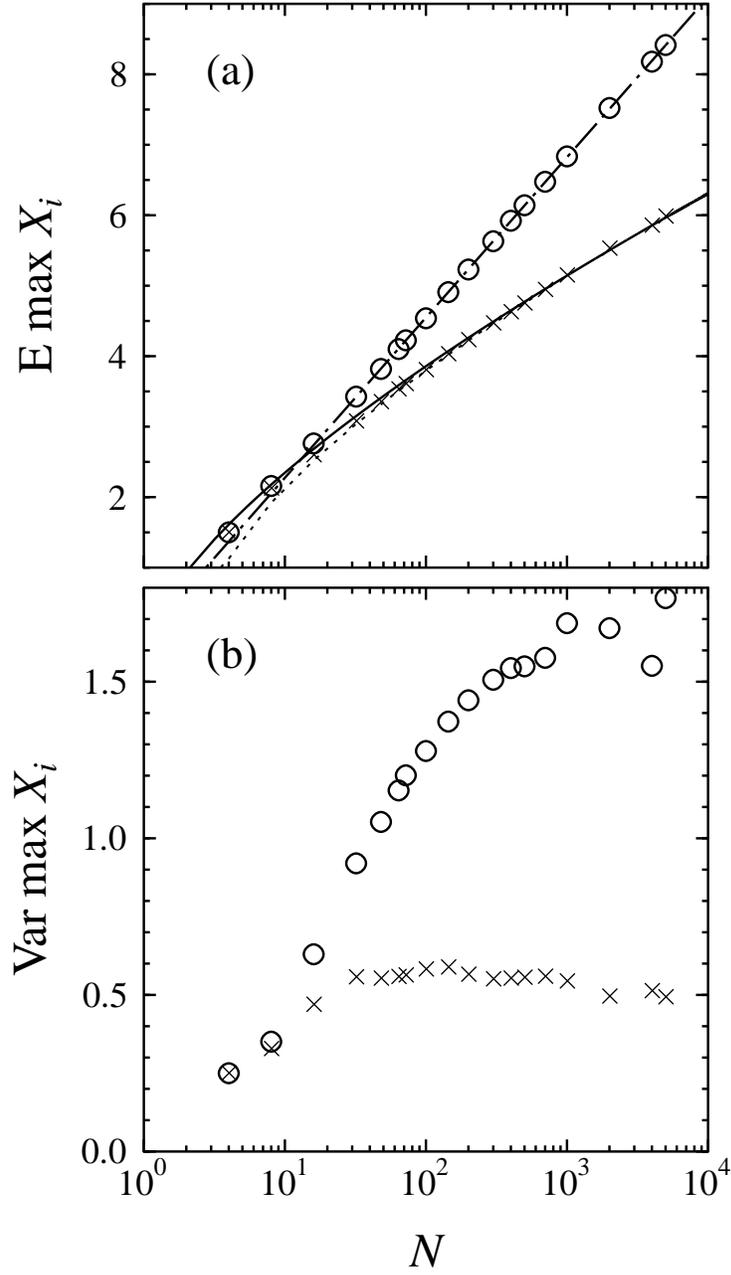}
        \caption{
          (a) Expectation values of the size of the longest particle 
          cluster and the size of the longest hole cluster and 
          (b) their variances as a function of the system size $N$ 
          for $z=n$ with $\rho=1/2$. 
          Crosses and squares denote the simulation data for 
          particle clusters and hole clusters, respectively. 
          In (a), the dotted curve was obtained by substituting 
          Eq.~(\ref{ASYCDFContSum}) in Eq.~(\ref{ASEPEMaxX}) 
          and the full curve by taking the crossover mass to be 
          the point where the approximate c.d.f.\ of 
          Eq.~(\ref{ASYCDFCont}) obtains the value 1/2. 
          The dash-dotted line is 
          the function $0.987\log N$, where the prefactor was chosen 
          to produce the 'best fit'. 
          \protect\label{FigASY}} 
\end{figure}

Again, one can show that the cluster sizes are asymptotically independent, 
in that Eq.~(\ref{AsymptoticIID}) holds with 
\begin{equation}
P(M_i =k) = {\frac{(N-n-1)^{n-k}}{(N-n)^n}}{n \choose k} 
 \longrightarrow
P_\infty (M_i =k) = {\frac{\lambda^k e^{-\lambda}}{k!}} 
\ \ {\textrm{as }} N \to \infty.
\end{equation}
Therefore, Eq.~(\ref{ASYCDFTail}) suggests that the 
limiting distribution 
function is, as in the $z=0$ case, approximately given by the product  
\begin{eqnarray}
F_{\lambda,n}(k) &\sim& \left( 1 - e^{-\lambda}\sum_{j\geq k+1} 
{\frac{\lambda^j}{j!}} \right)^{n/\lambda} \nonumber\\
\label{ASYCDFContSum}
&\approx& e^{-{\frac{e^{-\lambda}n}{\lambda}}\sum_{j\geq k+1} 
{\frac{\lambda^j}{j!}}} \\
\label{ASYCDFCont}
&\approx& e^{-{\frac{e^{-\lambda}\lambda^k n}{(k+1)!}}}, 
\end{eqnarray}
where $k > \lambda$ and $N$ is large. 
From the mathematical point of view, however, 
one should be careful with this approximation, 
because the distribution function for the maximum of i.i.d.\ Poisson 
variables does not actually converge to the Gumbel distribution - not 
even in sense of continuous Gumbel envelopes as it was in the 
$z=0$ case {\cite{{Anderson70},{Kimber83}}}. 
Namely, Anderson has shown {\cite{Anderson70}} that 
the probability of the maximum concentrates on two consecutive integers, 
i.e.\ there exists a sequence of integers $I_n(\lambda)$ such that 
\begin{equation}
\label{ASYLimit}
\lim_{n/\lambda \to \infty} 
P(\max_{1 \leq i \leq n/\lambda} M_i =I_n(\lambda)\ 
{\textrm{or}}\ I_{n}(\lambda) +1) = 1.
\end{equation}
The sequence $(I_n)$ then obviously determines the type of 
divergence of the mean. 
It was shown by Kimber {\cite{Kimber83}} that 
in the leading order this sequence diverges as 
\begin{equation}
\label{I_n}
I_n(\lambda) \sim {\frac{\log n}{\log \log n}}. 
\end{equation}
However, we would like to remark that, despite of its shortcomings, 
the approximation of Eq.~(\ref{ASYCDFCont}) yields the same 
functional form for the 
expected length of the longest cluster: In the leading order   
\begin{equation}
\langle\max_i X_i\rangle \approx \sum_{k=0}^{n-1} 
\chi_{\lbrace k \leq m_x \rbrace},
\end{equation}
where $\chi_{\lbrace \cdot \rbrace}$ is the indicator 
function and $m_x$ is some crossover mass, where the distribution 
function increases most rapidly. Eq.~(\ref{ASYCDFCont}) then 
gives the very interesting result
\begin{equation}
\label{ASYEDiv}
{\frac{(m_x +1)!}{\lambda^{m_x}}} \sim e^{-\lambda} n,
\end{equation}
i.e.\ the expected length of the longest {\it particle} cluster 
diverges as the inverse relation of Eq.~(\ref{ASYEDiv}) in $n$. 
In the special case $\lambda = 1$, i.e.\ 
$\rho = 1/2$, the type of divergence reduces to inverse factorial. 
Furthermore, by Stirling's formula one obtains the same functional form 
as in Kimber's result in Eq.~(\ref{I_n}). 
Our analytical result is again supported by simulations 
as seen in Fig.~\ref{FigASY}(a).

The plot of the variance in Fig.~\ref{FigASY}(b) shows again 
fluctuations, and the divergence of the distance between maxima is 
now dictated by the inverse relation of Eq.~(\ref{ASYEDiv}). 
In general, it can be seen that the fluctuations are more 
pronounced than those in the ordinary TASEP.  
We also observe that the variance has a global maximum at a 
finite system size. Below this maximum, 
the correlation length $\lambda$ is comparable to the system size, which 
results in large fluctuations of the maximum. For $N \gg \lambda$ the system 
can be considered to consist of a large number of independent copies 
and the i.i.d.\ behavior is recovered. In the end, one should notice that, 
unfortunately, the continuum approximation of the form 
Eq.~(\ref{ASYCDFCont}) with the factorial replaced by the gamma function 
cannot be used to evaluate the variance: In the continuous case, the maximum 
concentrates on one real number, in that the properly 
scaled variable converges to a 
distribution degenerate at zero \cite{{Galambos78},{Anderson70}}.

In Fig.~\ref{FigASY} we show the expectation and the variance also 
for the {\it hole} clusters. 
According to Eq.~(\ref{HoleDistr}), the limiting distribution for the 
lengths of the hole clusters is again geometric, as in the case of $z=0$ 
dynamics, but the number of clusters is random and 
the length of the longest cluster depends on that number. 
The solution to this problem can be found by defining new mass 
variables $M_i^*$ as the number of holes between $i^{\mathrm{th}}$ 
and $(i+1)^{\mathrm{th}}$ particle. Eq.~(\ref{HoleCDF}) then tells that
\begin{equation}
P(M_i^* \leq k) = 1 - C_{\rho,N} 
\left( 1 - {\frac{k}{N(1-\rho)}} \right)^{\rho N} 
\longrightarrow 1 -  C_{\rho}e^{-\lambda k} \qquad 
{\textrm{as }} N \to \infty,
\end{equation}
where $(C_{\rho,N})$ is a sequence of constants with 
$\lim C_{\rho,N} = C_{\rho} > 0$.
Since there are now $n$ new mass variables and the parent distribution is 
geometric-like, the limiting extremal distribution has again Gumbel 
envelopes and the divergence of the mean is logarithmic, 
as seen in Fig.~\ref{FigASY}(a).

\section{Discussion}
\label{SecDisc}

We shall now discuss our results from a different point of view.
There is a well-known mapping from TASEP to a certain lattice model, 
namely the one-dimensional single-step model, of interface 
growth \cite{DerridaPrivman97}. 
In this mapping each particle is considered a unit step down 
and each vacancy a unit step up. 
The function $h(x,t)$, where $x$ is the 
spatial and $t$ the time coordinate, obtained this way 
is defined to be constant between the integers and, say, right-continuous. 
Clearly, the condition $\rho \neq 1/2$ implies existence of a global tilt.    
The lengths of the clusters in the TASEP picture translate to the 
lengths of the decreasing parts of the interface. 
Note that the $z=0$ case has particle-hole symmetry 
and the results obtained for the decreasing parts (corresponding to particle
clusters) are valid for the increasing parts (hole clusters) as well. 
For $z \geq 1$ the properties of the increasing parts must be 
calculated from the hole distribution.

\begin{figure}[htb]
\epsfxsize=\columnwidth\epsfbox{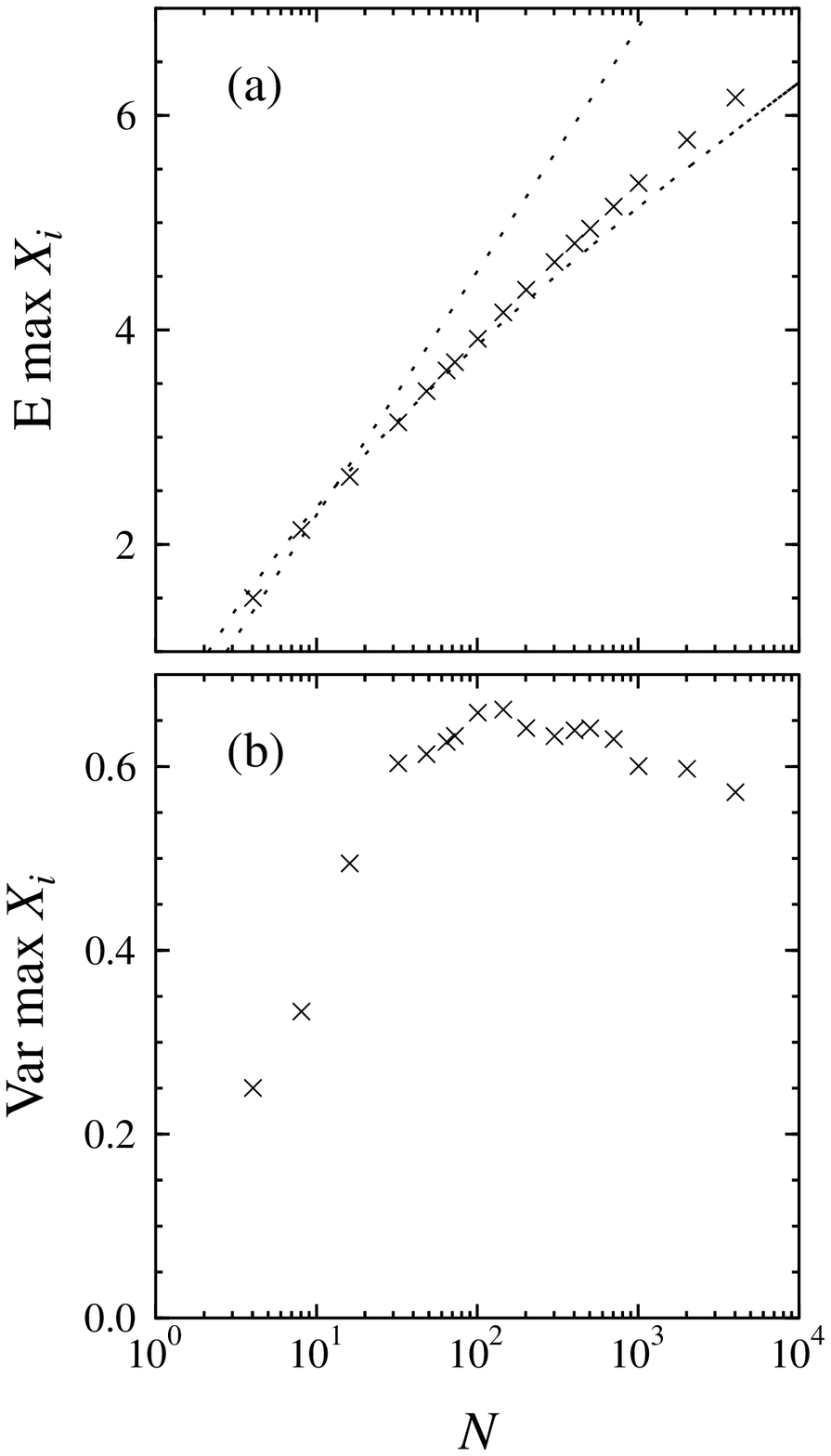}
      \caption{
        (a) Expectation values of the size of the longest cluster and
        (b) its variance as a function of the system size $N$ 
        for the symmetrized model with $z=n$ and $\rho=1/2$. 
        Crosses denote the simulation data for clusters in the 
        symmetrized model. 
        In (a), the asymptotic curves for the corresponding asymmetric 
        case from Fig.~\ref{FigASY} are shown for comparison by the 
        dashed lines. 
      \protect\label{FigSYM}}
\end{figure}

The $z \geq 1$ dynamics can be seen to be equivalent to deposition rule 
such that the deposited particles relax towards the local minima from the 
decreasing parts of the height function $h(x,t)$. 
If the particle lands on an increasing part, nothing happens. 
Since this kind of asymmetry is not very common in real applications 
of the interface interpretation, we also studied numerically a spatial 
symmetrization of the $z=n$ case with $\rho = 1/2$.
This symmetrization belongs to the Edwards-Wilkinson 
universality class \cite{MeakinKirja}. 
In it, deposited particles always flow downhill to 
a local minimum. In case the particle lands on a hill top, the minimum 
is determined using a fair coin. 
The results are plotted in Fig.~\ref{FigSYM}. 
The mean length of the longest cluster diverges now faster than in 
the asymmetric $z=n$ case studied in Sec.~\ref{SecASY}, the earlier
analytical result for which is shown for comparison in Fig.~\ref{FigSYM}(a). 
This is consistent with the fact that on average 
the hole clusters are considerably 
shorter than in the case of asymmetric dynamics, which is compensated by 
longer particle clusters. 
The other features can be seen to be quite similar to those of the  
asymmetric case. For example, the variance of the longest cluster 
has a maximum at finite $N$ and quasiperiodic oscillations as before.

In conclusion, we presented analytical and numerical results for the 
properties of the clusters in asymmetric exclusion processes with two 
different dynamics. In particular, we studied the properties of the 
longest cluster in the system and showed that the large system behavior 
agrees with the asymptotic theory of the extremes for independent and 
identically distributed variables. The expected length of the longest cluster 
was found to diverge logarithmically with increasing system size for the 
ordinary TASEP and as logarithm divided by double logarithm in the case of 
modified dynamics, which corresponds to independent random walkers in the 
mass picture. In the latter case, the length of the longest cluster is 
distributed among two consecutive integers in the large 
system limit.

\bigskip

\section*{Acknowledgements}

We thank Dr.~Vesa Ruuska and Prof.~Stefan Geiss for discussions. 
This work has been supported by the Academy of Finland under 
the Center of Excellence Program (Project No.~44875).

\begin{thebibliography}{100}

\bibitem[\ast]{EmailOtto}
Corresponding author, Email address: otto.pulkkinen@phys.jyu.fi

\bibitem[\dag]{EmailJuha}
Email address: juha.merikoski@phys.jyu.fi

\bibitem{DerridaPrivman97}
B. Derrida and M. R. Evans, 
in {\it Nonequilibrium statistical mechanics in one Dimension},
Edited by V. Privman (Cambridge University Press, Cambridge, 1997).

\bibitem{Liggett1}
T. M. Liggett, {\it Interacting particle systems} 
(Springer, Berlin, 1985).

\bibitem{Liggett2}
T. M. Liggett, {\it Stochastic interacting systems: contact, voter and 
exclusion processes} 
(Springer, Berlin, 1999).

\bibitem{Derrida93_1}
B. Derrida, M. R. Evans, V. Hakim, and V. Pasquier, J. Phys A {\bf 26}, 
1493 (1993). 

\bibitem{Derrida93_2}
B. Derrida, M. R. Evans, and D. Mukamel, J. Phys A {\bf 26}, 
4911 (1993).

\bibitem{Derrida_cond-mat01}
B. Derrida, J. L. Lebowitz, and E. R. Speer, cond-mat/0105110.

\bibitem{Schadschneider93}
A. Schadschneider and M. Schreckenberg, J. Phys A {\bf 26}, L679 (1993).

\bibitem{Nagatani93}
T. Nagatani, J. Phys A {\bf 26}, 6625 (1993).

\bibitem{Bazant00}
M. Z. Bazant, Phys. Rev. E {\bf 62}, 1660 (2000).

\bibitem{Evans00}
M. R. Evans, Brazilian J. of Phys. {\bf 30}, 42 (2000), 
cond-mat/0007293. 

\bibitem{Majumdar99}
S. N. Majumdar, S. Krishnamurthy and M. Barma, 
J. Stat. Phys. {\bf 99}, 1 (2000). 

\bibitem{Spitzer70}
F. Spitzer, Adv. Math. {\bf 5}, 246 (1970). 

\bibitem{Bremaud99}
P. Br\'emaud, {\it Markov chains: Gibbs fields, Monte Carlo simulation, 
and queues} (Springer, New York, 1999).

\bibitem{Galambos78}
J. Galambos, {\it The asymptotic theory of extreme order statistics} 
(Wiley, New York, 1978).

%\bibitem{Galambos88}
%J. Galambos, {\it Advanced probability theory} (Dekker, New York, 1988).

\bibitem{KotzKirja}
S. Kotz and S. Nadarajah, {\it Extreme value distributions: 
theory and applications} (Imperial College Press, London, 2000).

\bibitem{Anderson70}
C.W. Anderson, J. Appl. Prob. {\bf 7}, 99 (1970). 

\bibitem{Simulation}
All our simulation data was obtained by a standard Monte 
Carlo procedure using the random number generator \textsc{ranmar}. 

\bibitem{Kimber83}
A. C. Kimber, Z. Wahrscheinlichkeitstheorie verw. Gebiete {\bf 63}, 
551 (1980).

\bibitem{MeakinKirja}
P. Meakin, {\it Fractals, scaling and growth far from equilibrium} 
(Cambridge University Press, Cambridge, 1998).


\end{thebibliography}
\end{document}